\documentstyle[11pt,paspconf,epsf]{article}

\begin{document}

\title{Halo evolution in a cosmological environment}

\author{S. Gottl\"ober \altaffilmark{1}}
\affil{Astrophysical Institute Potsdam, An der Sternwarte 16, D-14482
Potsdam, Germany}

\author{A.A. Klypin, A.V. Kravtsov}
\affil{Astronomy Department, NMSU, Dept.4500, Las Cruces, NM
88003-0001, USA}

\altaffiltext{1}{Also: Astronomy Department, NMSU, Dept.4500, Las Cruces,
NM 88003-0001}

\begin{abstract}

We present results of a study of the formation and evolution of the
dark matter (DM) halos in a {\sl COBE}-normalized spatially flat
$\Lambda$CDM model ($\Omega_0=1-\Omega_{\Lambda}=0.3$; $h=0.7$).  The
dynamics of $256^3$ DM particles is followed numerically in a box of
$60 h^{-1}$ Mpc with the dynamic range of $32,000$ in spatial
resolution. The high resolution of the simulation allows us to examine
evolution of both {\em isolated} and {\em satellite} halos in a
representative volume. We discuss the new halo finding algorithm
designed to identify halos in high-density environments,
present results on the evolution of velocity function of DM halos and
compare it with the Press-Schechter function, discuss the evolution of
power spectrum of matter and halo distributions, and mass evolution of
halos. 

The velocity function of halos at $z=0$ compares well with the
prediction of the Press-Schechter approximation, but for circular
velocities in the range 100 -- 200 km/s simulations predict $\sim 1.3$
time more halos (mostly in clusters or groups).
In real space the power spectra of halos and DM are very different
(halos are anti-biased). Both spectra do not have simple power-law
shape. In redshift space the spectra are close to a power law with
$\gamma = -2.1$ in the range of wave numbers $k = 0.2 - 5 h
$Mpc$^{-1}$.  The power spectra of halo distribution evolves only
mildly for $z= 0 - 3$.
The mass evolution of {\it isolated} virialized objects determined from
the simulation is in good agreement with predictions of the extended
Press-Schechter models.  However, satellite halos evolve very
different: for some of them the mass decreases with time, which happens
if the halos fall into clusters or groups. We discuss the dependence of
the correlation function of halo populations on their environment and
merging history.

\end{abstract}



\section{Introduction}
A variety of observations on a wide range of spatial scales
(i.e. rotation curves of galaxies, galaxy dynamics and gravitational
lensing in galaxy clusters, the large-scale velocity flows, etc.)
indicate the existence of dark matter (DM) in the Universe.  The total
amount of DM is not yet known, but it is generally
believed that cold dark matter dominates the mass in the Universe and
significantly affects both the process of galaxy formation and the
large-scale distribution of galaxies. The gravitational domination of
DM on the scale of galactic virial radius implies that collisionless
simulations can be used to study formation of the DM component of
galaxies.

It is well known that cosmological scenarios with cold dark matter
alone cannot explain the structure formation both on small and very
large scales.  Recently, scenarios with a non-zero vacuum energy,
quantified by the cosmological constant $\Lambda$, have been proved to
be very successful in describing most of the observational data at both
low and high redshifts.  In this contribution, we discuss the evolution
of DM halos in a spatially flat cosmological model dominated by the
cosmological constant and cold dark matter, $\Lambda$CDM. A simulated
box of 60 $h^{-1}$ Mpc size allows to construct and study statistically
representative halo catalogs (about 10,000 halos are identified at
$z=0$). The novel feature of this analysis is study of both isolated
and satellite halos, where halos are dubbed satellites if they are
located within the virial radius of a larger host system (massive
galaxies or galaxy groups and clusters).  This allows us to study the
differences between isolated and satellite halos; here we discuss the
differences in their mass evolution and spatial distribution.

\section{Numerical simulation}

We simulate the evolution of $256^3$ cold dark matter particles in a
spatially flat model with a cosmological constant ($\Lambda$CDM; 
$\Omega_0=1-\Omega_{\Lambda}=0.3$; $\sigma_8=1.0$; $H_0=70$ km/s/Mpc). 
The model is normalized in accord with the four year {\sl COBE} DMR
observations (Bunn \& White 1997) and observed abundance of galaxy clusters
(Viana \& Liddle 1996).  The age of the universe in this model is
$\approx 13.5$ Gyrs.

In order to study the statistical properties of halos in a cosmological
environment, the simulation box should be sufficiently large. On the
other hand, to assure that halo survival in clusters, the force
resolution should be ${_ <\atop{^\sim}} 1-3h^{-1} {\rm kpc}$ and the
mass resolution should be ${_ <\atop{^\sim}} 10^9h^{-1} {\rm M_{\odot}}$
(Moore et al. 1996, Klypin et al. 1998, hereafter KGKK).  The
simulations were done using the Adaptive Refinement Tree (ART) code
(Kravtsov, Klypin \& Khokhlov 1997). The code used a $512^3$
homogeneous grid on the lowest level of resolution and six levels of
refinement, each successive refinement level doubling the
resolution. The sixth refinement level corresponds to a formal
dynamical range of 32,000 in high density regions. Thus we can reach in
a 60 $h^{-1}$ Mpc box the force resolution of $\approx 2h^{-1}$
kpc. The simulation has a mass resolution (particle mass of $1.1 \times
10^9 h^{-1} {\rm M_{\odot}}$) sufficient to identify galaxy-size halos
($M {_ >\atop{^\sim}} 3 \times 10^{10}h^{-1} {\rm M_{\odot}}$).

The ART code integrates the equations of motion in {\em comoving}
coordinates.  However, its refinement strategy is designed to
effectively preserve the initial {\em physical} resolution of the
simulation. In order to prevent degradation of force resolution in {\em
physical} coordinates, the dynamic range between the start ($z_i = 30$)
and the end of the simulation should increase by $(1+z_i)$: For our
simulations it should have at least the dynamical range
$512\times(1+z_i)=15,872$.  This is accomplished with the prompt
successive refinements {\em in high-density regions} during the
simulations. The peak resolution is reached by creating a hierarchy
with six levels of refinement. The spatial refinement is accompanied by
the similar refinement of the integration time step.  The integration
step of $\Delta a_0=0.0015$ corresponds to 645 time steps on the
lowest-resolution of the uniform grid and to (effective) 41,280 on the
deepest refinement level .  In physical units, the latter step
corresponds to $2.3 \times 10^5$ years.

\section{Challenges in halo identification}
Identification of halos in dense environments and reconstruction of
their evolution is a challenge. Any halo finding algorithm has to deal
with difficult ``decision-making'' situations, existing also in the
real Universe. The most typical difficulties arise if a small
satellite is bound to a larger galaxy (like the LMC and the Milky Way
or the M51 system) or in cases when many small gravitationally
self-bound halos are moving within a large region of virial
overdensity (galaxy clusters and groups).

Assuming that the satellite is self-gravitationally bound, we would
have to include the mass of the satellite in the mass of the host
system. By doing so, we count mass of the satellite twice: when we
identify the satellite and when we identify the host system. This may
seem unreasonable, but if we do not include the satellite, then the
mass of the large galaxy is underestimated. For example, the binding
energy of a particle at the distance of the satellite will be
wrong. The problems arise also when we try to assign particles to
different halos in the effort to find halo masses. This is very
difficult to do for particles moving between halos. Even if a particle
at some moment is bound to one of the halos, it is not guaranteed that
it belongs to the halo. The gravitational potential changes with time,
and the particle may end up falling onto another halo. These effects
are especially frequent in high-density crowded regions typical of
galaxy groups and clusters. This situation actually was observed very
often in simulated halos when we compared particle contents of halos at
different redshifts.  It represents an additional difficulty in tracing
the mass history of such halos.

It is also difficult to estimate the actual mass of a halo orbiting
inside a cluster-size object. The formal virial radius of such a
satellite halo is simply equal to the cluster's virial radius.  We have
chosen to define the outer {\em tidal} radius of the satellite halos as
a scale at which their density profile starts to flatten. At small
distances from the center of the satellite the density steeply
declines, but then it flattens out and may even increase. This means
that we reached the outer border of the satellite. The corresponding
mass within tidal radius is then our approximation to the halo mass.
With these considerations in mind, the halo finding algorithm must
allow halos to overlap and DM particles to belong to more than one
halo.

\section{Halo finding algorithm}
The most widely used halo-finding algorithms, the friends-of-friends (FOF)
and the spherical overdensity, both discard ``halos inside halos'',
i.e., satellite halos located within the virial radius of larger halos. The
distribution of halos identified in this way, cannot be compared easily
to the distribution of galaxies, because the latter are found within
larger systems. In order to cure this, we have developed two related
algorithms, which we called the {\em hierarchical friends-of-friends}
(HFOF) and the {\em bound density maxima} (BDM) algorithms (KGKK). 

Since the algorithms work on a snapshot of the particle distribution,
they tend to identify also small fake ``halos'' consisting of only a
few {\em unbound} particles, clumped together by chance at the analyzed
moment. We deal with this problem by both checking whether the
identified clump is gravitationally bound and by following the merging
history of halos. A halo that does not have a progenitor at the previous
moment is discarded, if the particles from which this halo has been
formed belong at the previous moment already to other halos. In this
case the halo is assumed to be a fake one. If it has formed from single
particles or small objects below the threshold of halo detection, it is
assumed that a new halo has been formed. For other halos we find the
direct progenitor, i.e.  a halo at a previous moment that contains the
maximum number of particles of this halo. We use the chain of
progenitors identified in this way to reconstruct the mass evolution of
the given halo back in time, down to the epoch of its first detection
in the simulation.

The HFOF and BDM algorithms are complementary. Both of them find
essentially the same halos. Therefore, we believe that each of them is
a stable algorithm which finds in a given dark matter distribution all
dark matter halos above a given mass threshold.  The advantage of the HFOF
algorithm is that it can handle halos of arbitrary, not only
spherically symmetric, shape. The advantage of the BDM algorithm is
that it describes better the physical properties of the halos due to
the fact that it separates background unbound particles from the
particles gravitationally bound to the halo.

\section{Velocity function}

Interacting halos exchange mass and lose mass. The total mass of a halo
depends on its radius which, as was pointed out above, is difficult to
define. We try to avoid this problem by assigning not only a mass to a
halo, but finding also its maximum ``circular velocity''
($\sqrt{GM/R}$), $v_{circ}$. This is the quantity which is more
meaningful observationally. Numerically, $v_{circ}$ can be measured
more easily and more accurately then mass.

The output of the halo finding algorithm depends primarily on the
assumed mass threshold. With the threshold of $10^{10} h^{-1} {\rm
M_{\odot}}$ (10 particles) the algorithm identifies $\sim 17000$
halos, whereas a mass threshold of $3 \times 10^{10} h^{-1} {\rm
M_{\odot}}$ (28 particles) results in identification of more than 9000
halos. There is a weak dependence on the assumed maximum halo radius
(2\% decrease if changing the maximum radius from $100 h^{-1}$ kpc to
$150 h^{-1}$ kpc).

For any study one needs to have a complete halo sample that is not
affected by the numerical details of halo finding procedure. In order
to test the completeness of the halo samples, we have constructed the
differential velocity functions at $z=0$ for different mass thresholds
and maximum radii. We find that the halo samples do not depend on the
numerical parameters of the halo finder for halos with $v_{circ} 
{_ >\atop{^\sim}} 100$ km/s.  Particularly, the differential velocity
function for $v_{circ} > 100$ km/s is robust and does not depend on the
assumptions.  There is a substantial scatter for $v_{circ} > 500$ km/s
where we have less than 10 halos per logarithmic velocity bin. In fact,
increasing the maximum halo radius one (slightly) increases also the
maximum circular velocity, if it is not yet reached. This is the case
for the most massive halos. We reanalyzed these halos and found that in
all cases the maximum circular velocity changed by less than 10 \%.

\begin{figure}
\plottwo{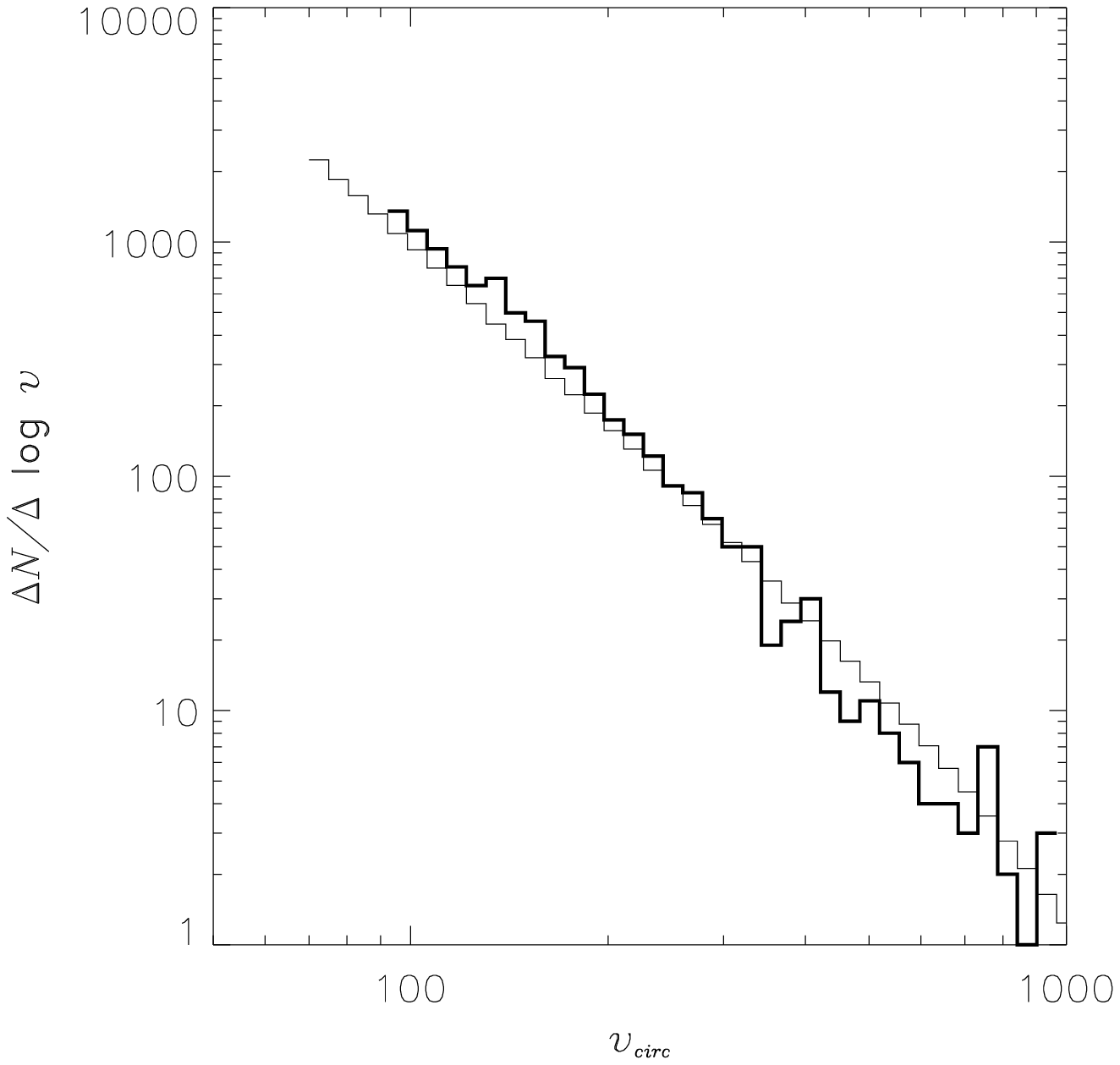}{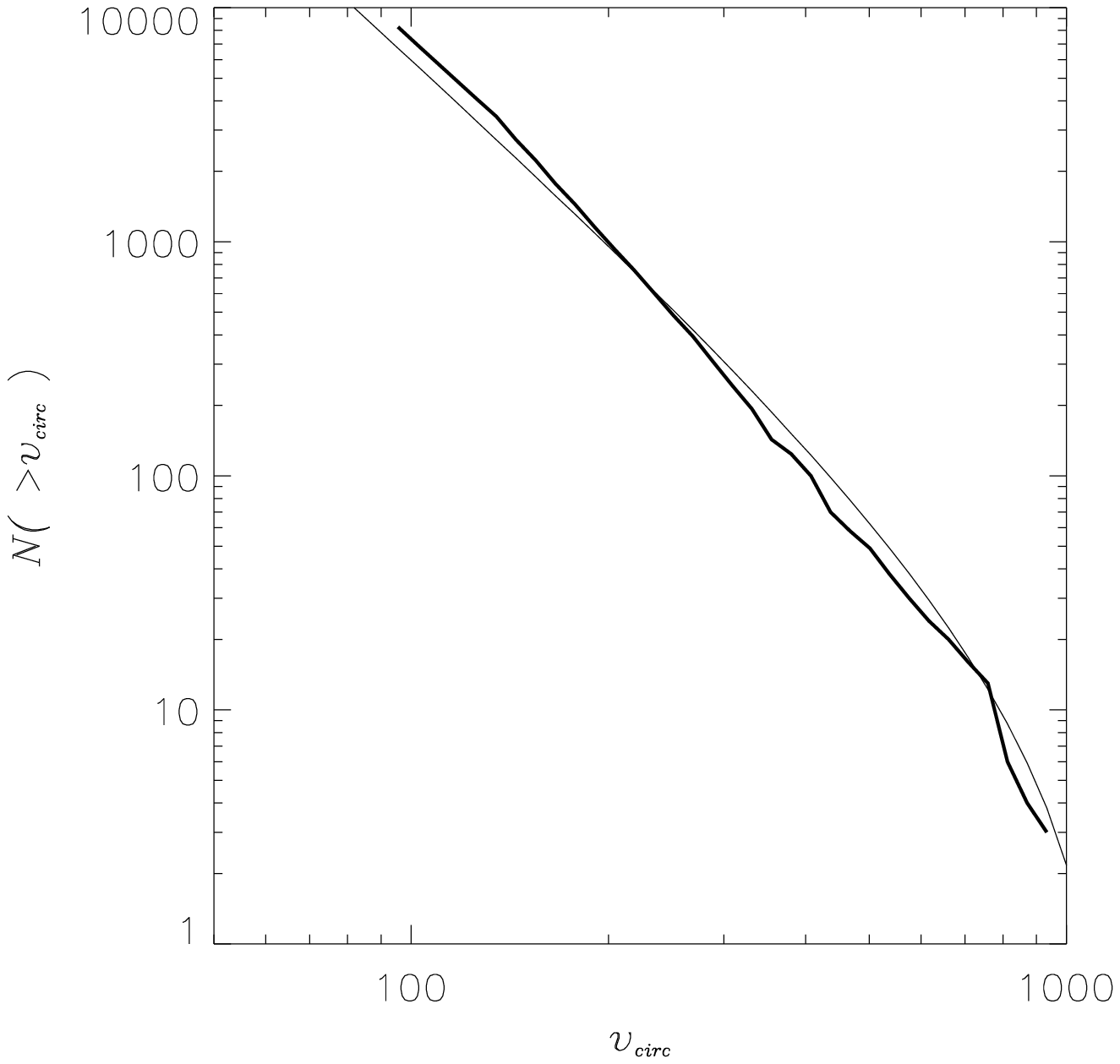}
\caption{Differential (left panel) and integral (right panel) 
velocity functions of halos at $z=0$. The
thick lines show the velocity function of halos, while thin lines
represent the Press-Schechter prediction.}
\label{vel_bin}
\end{figure}

In Figure~\ref{vel_bin}, we compare the velocity function of halos in
our simulation with the Press-Schechter (PS) prediction. The latter has
been been computed by converting virial mass in the PS mass function to
$v_{circ}$, assuming the NFW density profile of the halos. There is an
excess of small halos ($v_{circ} < 200$ km/s) over the PS prediction.
This is expected because we include satellite halos in our catalog,
while PS model predicts mass function of the isolated objects. For
higher circular velocities we find a slight deficit of massive halos
(20\% - 30 \%) in comparison with the numbers predicted by the PS mass
function. In particular, for the largest halos, the over-prediction may
be influenced by the assumption of NFW density profile which is a
rather poor description for some of the large halos with substantial
central sub-structure.  The prediction of the simulation is also
influenced by cosmic variance in our single realization; we may have a
deficit of massive halos in the given realization just by chance.

\section{Halo evolution}

\begin{figure}
\vspace*{5.0cm}
\includegraphics{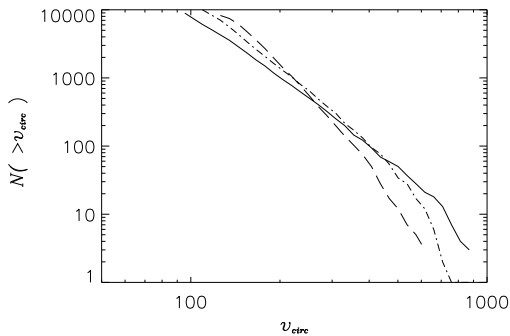}
\caption{Evolution of the integral velocity function of halos. The solid, 
dot-dashed, and dashed curves correspond to $z=0$, $z=1$, and $z=3$,
respectively.  }
\label{vel_z}
\end{figure}

In the following we will discuss first the evolution of halo clustering
and then the evolution of the halos. We consider three catalogs of
halos with $v_{circ}>120$ km/s at $z = 0, \;1, \;3$ (4804, 8018, 8396
halos respectively). These samples are are complete for $z = 0, \;1$:
the differential velocity function does not show any turnoff (see
Figure~\ref{vel_z}) at these velocities. We are missing part of the
halos with $v_{circ} < 130$ km/s at $z=3$.  To study the evolution of
the halos we construct the mass evolution history of all halos which
have been identified in the simulation at $z=0$.

\subsection{Evolution of halo clustering} \label{evol}

\begin{figure}
\plottwo{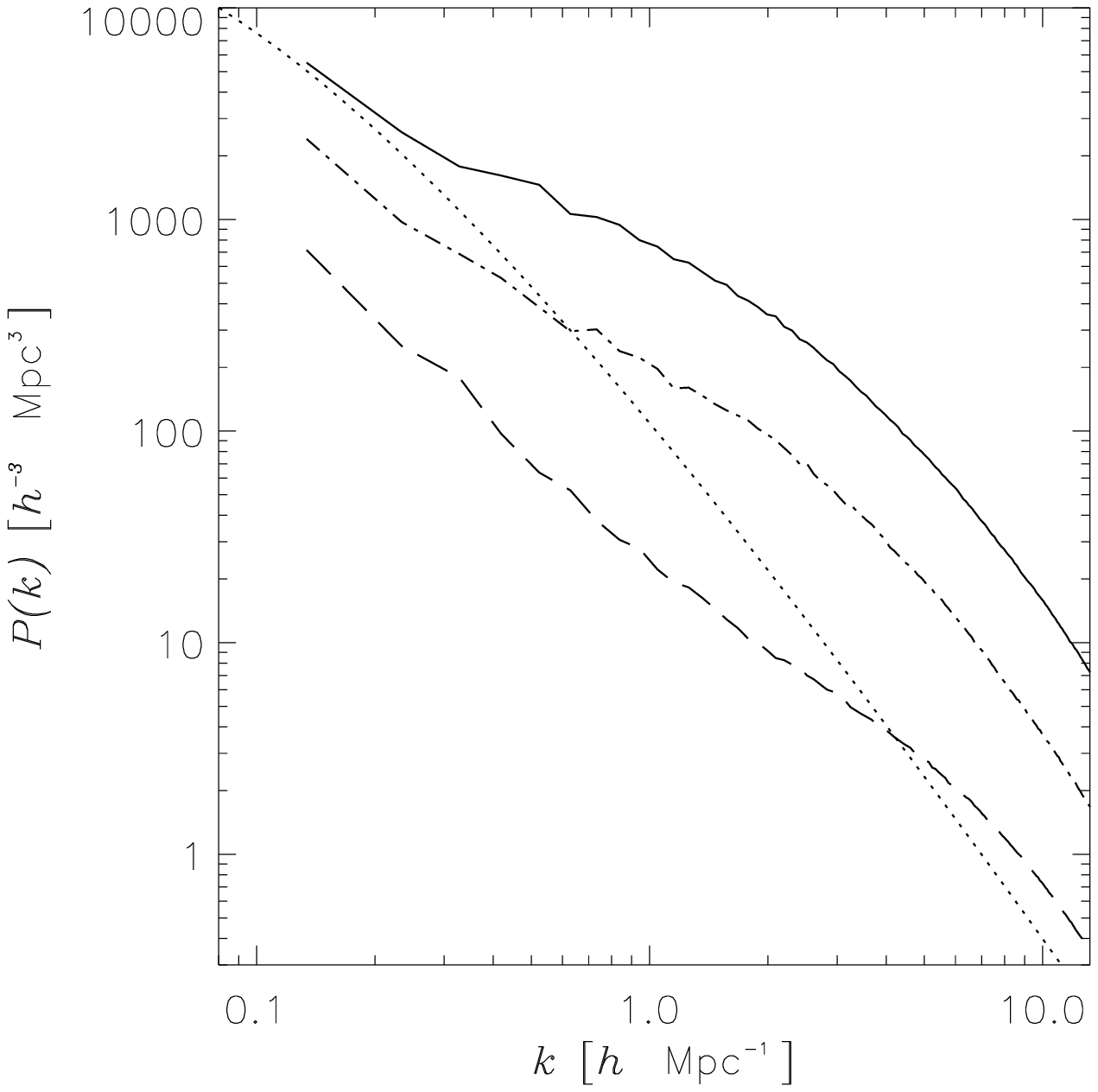}{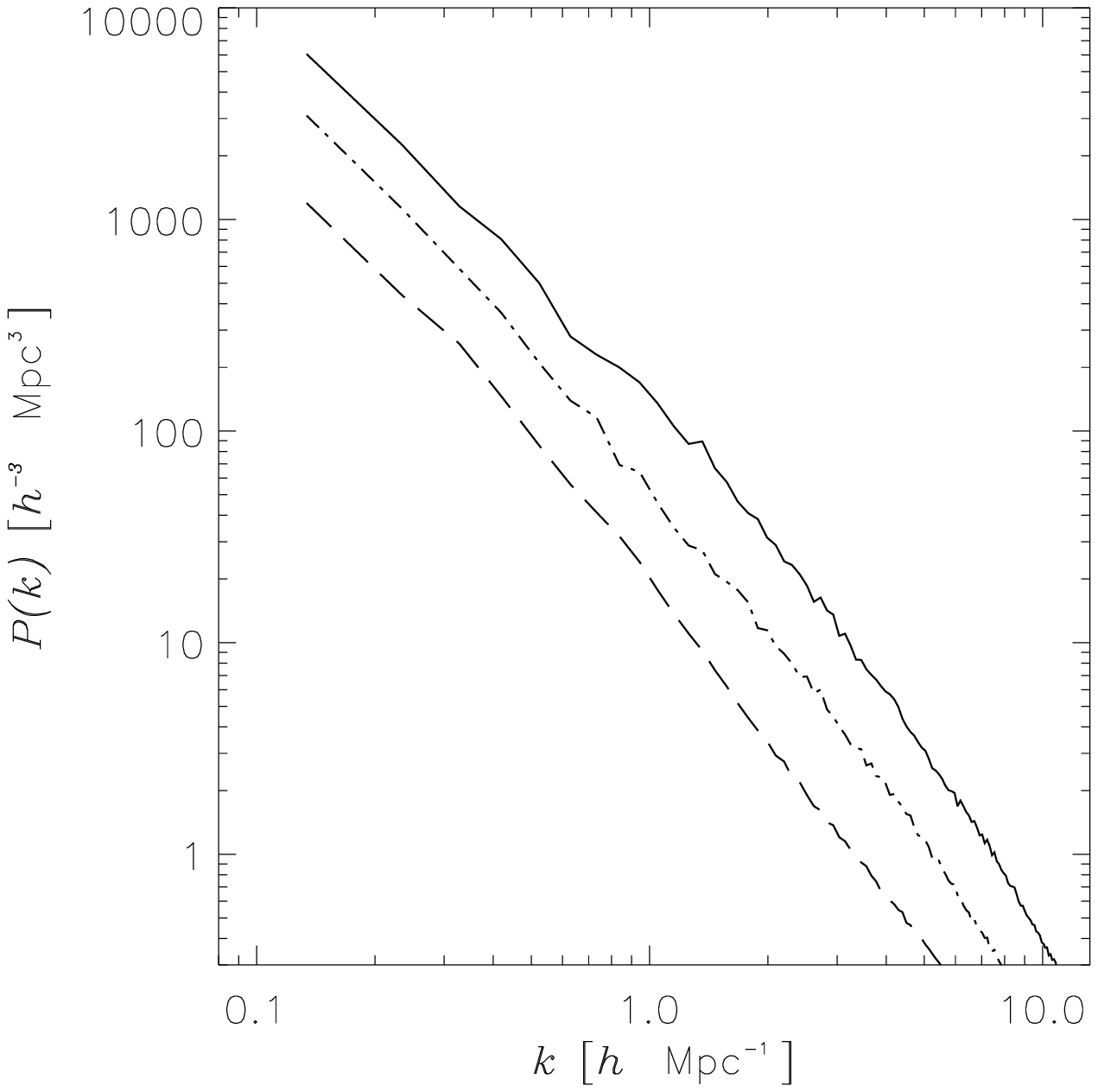}

\caption{Evolution of real-space (left) and redshift-space (right) 
power spectrum of the DM particles. The solid, dot-dashed, and dashed
curves correspond to $z=0$, $z=1$, and $z=3$, respectively.  The dotted
line left shows the linear power spectrum at $z = 0$.  }
\label{pk_dm}
\end{figure}

\begin{figure}
\plottwo{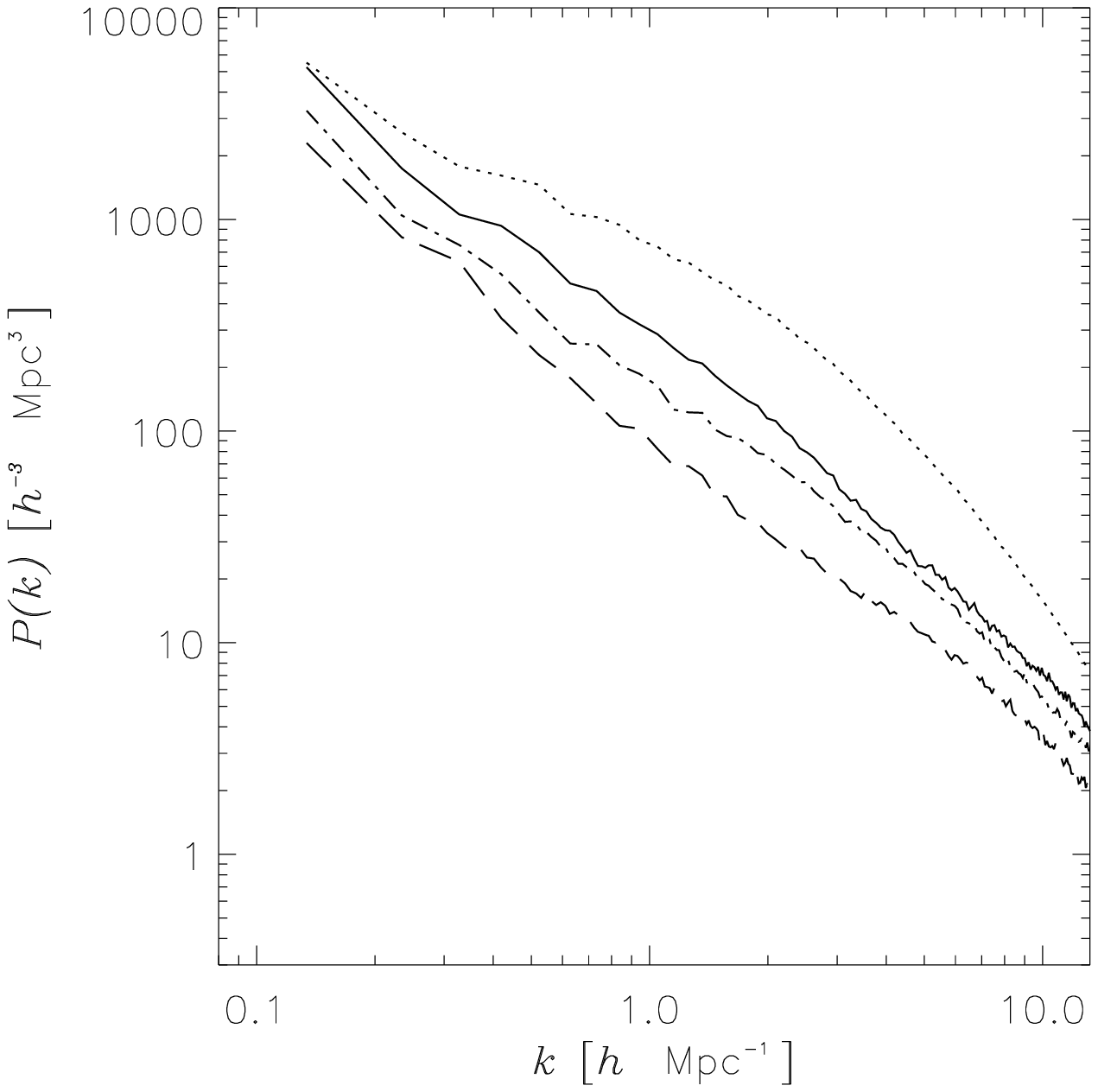}{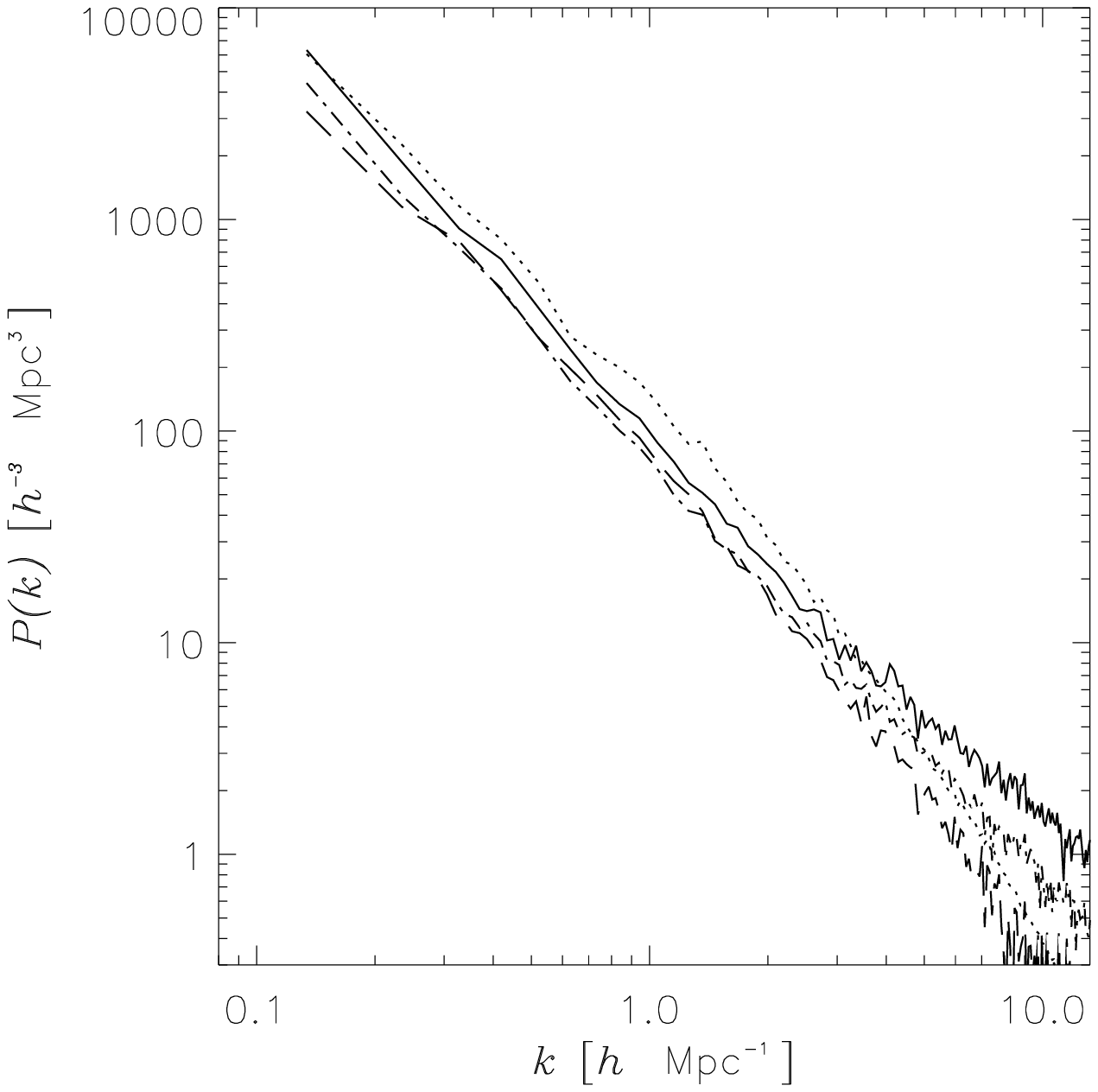}

\caption{Evolution of real-space (left) and redshift-space (right) 
power spectrum of halos with circular velocity $v_{circ} > 120$
km/s. The solid, dot-dashed, and dashed curves correspond to $z=0$,
$z=1$, and $z=3$, respectively. The dotted lines show the DM 
power spectrum at $z = 0$.  }
\label{pk_halo}
\end{figure}

In Figures~\ref{pk_dm} and \ref{pk_halo}, we show the evolution of the
power spectrum of these halos in comparison with the power spectrum of
the DM particles. We have calculated the power spectrum on a $512^3$
grid; the spectra are shown up to half of the Nyquist frequency
$k_{box} \times 128$. The shot noise level S = V/N for the halo samples
is 45, 27, and 26, respectively. The shot noise power derived from a
random distribution is constant up to about a quarter of the Nyquist
frequency $k_{box} \times 64$ and decrease than by about 25\%. We
expect an error of the same order in the halo power spectrum at
$k_{box} \times 128$.

The halo power spectrum evolves much slower than that of the dark
matter.  The halos are clearly biased at $z=3$, are essentially
unbiased at $z=1$, and are anti-biased at $z=0$. A detailed analysis of
the evolution of halo and matter power spectra and bias will be
presented elsewhere (Kravtsov and Klypin 1998, in preparation). The
results obtained from the power spectrum analysis are in qualitative
agreement with results on the bias evolution as derived from the
correlation function analysis by Col\'{\i}n et al. (1998) and other
researchers. The scale dependence of the bias, however, is quite
different.

In the right panels of Figures~\ref{pk_dm} and \ref{pk_halo}, we show
the power spectrum of halos and dark matter in {\em redshift} space.
The redshift space power spectra are always much steeper then in real
space. At $z=0$, the real space power spectrum of dark matter shows a
clear excess over the linear power spectrum due to nonlinear
clustering. The corresponding dark matter power spectrum in redshift
space follows almost exactly the linear power spectrum of dark matter.

In real space the power spectra of halos and DM are very
different. Both spectra do not have a simple power-law shape. The
real-space power spectrum of halo distribution evolves only mildly
between $z=3$ and $z=0$. Similarly to the real space case, the
redshift-space power spectrum of halos shows almost no evolution during
this time. In the range of wave numbers $k = 0.2 - 5 h$ Mpc$^{-1}$ the
redshift space power spectrum of halos is close to a power law with
$\gamma = -2.1$. A power spectrum slope of $\sim -2.1 $ had been
measured for the combined SSRS2 + CfA2 galaxy sample on scales 
${_ <\atop{^\sim}}  30 h^{-1}$ Mpc (da Costa et al. 1994).

\subsection{Progenitors of halos}

In order to study the evolution of individual halos, we need to
construct a complete evolution tree for each of the halos in the $z=0$
catalog. We proceed as follows: We have selected 19 epochs ($z = 0$,
0.05, 0.1, 0.2, 0.3, 0.4, 0.5, 0.6, 0.7, 0.8, 0.9, 1.0, 1.5, 2., 2.5,
3., 5., 10., 15.). For every epoch, we identify a progenitor of a $z=0$
halo. The procedure of progenitor identification is based on the
comparison of lists of particles belonging to the halos at different
moments both back and forward in time. As was mentioned above, the halo
finder algorithm allows halos to overlap, or, in other words allows
particles to belong to more than one halo.  The visual inspection of a
large number of constructed evolution trees showed that this
forward-backward algorithm of tracing halo histories identifies the
correct ``ancestor-descendant'' relationships rather accurately, with
obvious ancestor-descendant misidentifications in ${_ <\atop{^\sim}}
2\%$ of the cases.

\begin{figure}
\plottwo{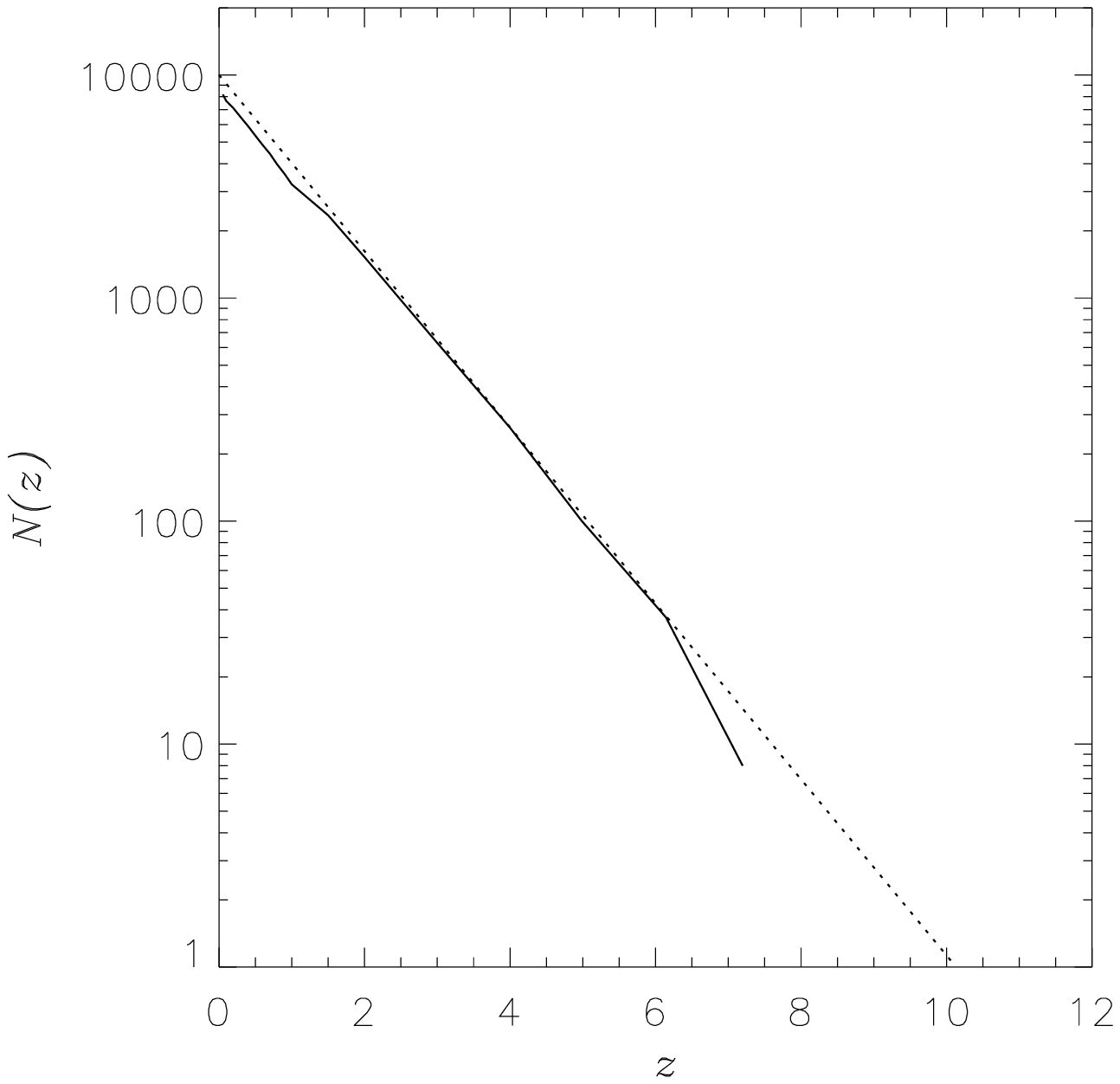}{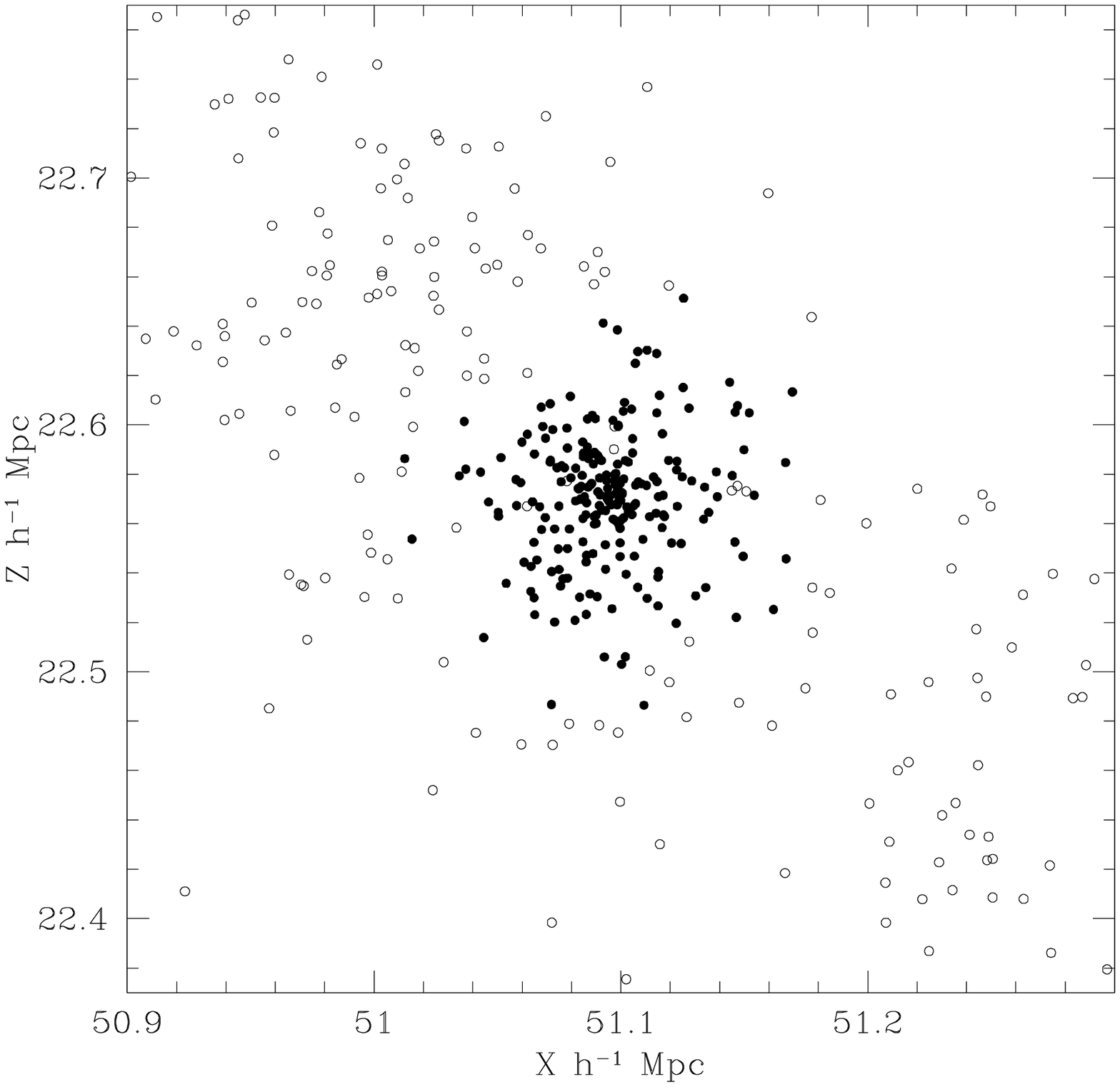}
\caption{Left: Number of halos in a $60 h^{-1} $Mpc simulation box
whose progenitors existed at redshift $z$. Only halos with mass larger
than $3 \times 10^{10}h^{-1} {\rm M_{\odot}}$ have been taken into
account. Dotted line: $ \propto\exp(-z/z_1),\;\;z_1 = 1.1$.  Right: A
virialized halo ($M\approx 1.95 \times 10^{11}{\rm M_{\odot}}$) at
$z=10$; filled circles show particles bound to the halo, the open
circles show unbound particles.}
\label{form_time}
\end{figure}

Using the procedure described above, we are able to address the
question of the {\em halo detection epoch}, which we define as an epoch
at which the halo has been identified for the first time.  The epoch of
the first identification depends, of course, on the mass threshold
assumed by the halo finder. The following results are for the lowest
possible mass $3 \times 10^{10}h^{-1} {\rm M_{\odot}}$ at $z=0$. For
this mass threshold the halo catalog is dominated by halos of mass
${_ <\atop{^\sim}}  10^{11}h^{-1} {\rm M_{\odot}}$, which fall below the
identification threshold quickly as we trace their mass evolution back
in time. The distribution of detection time for these halos is thus
likely to reflect selection function determined by the threshold rather
than any kind of physical ``formation epoch'' distribution. In the left
panel of Figure~\ref{form_time}, we show the distribution of detection
epochs for the halos identified with mass threshold of $3 \times
10^{10}h^{-1} {\rm M_{\odot}}$. The number of progenitors at a given
redshift exponentially decreases with redshift.  The figure shows that
the first halo in this catalog should have been detected as early as
$z=10$. Indeed, at $z=10$ we find one halo of mass $ 1.95 \times
10^{11}{\rm M_{\odot}}$ (184 bound particles shown as filled circles in
the right panel of Figure~\ref{form_time}). The overdensity of the halo
is 400 which means that it is a virialized object. We show also the
surrounding, not yet bound, particles of the halo (open circles). The
overdensity in the box of $400 h^{-1}$ kpc size centered on this object
is $\approx 50$.  The figure also shows a flattened structure of the
particle distribution: the halo forms inside of a small pancake.

\subsection{Halo mass evolution and environment} \label{masevol}

\begin{figure}
\plottwo{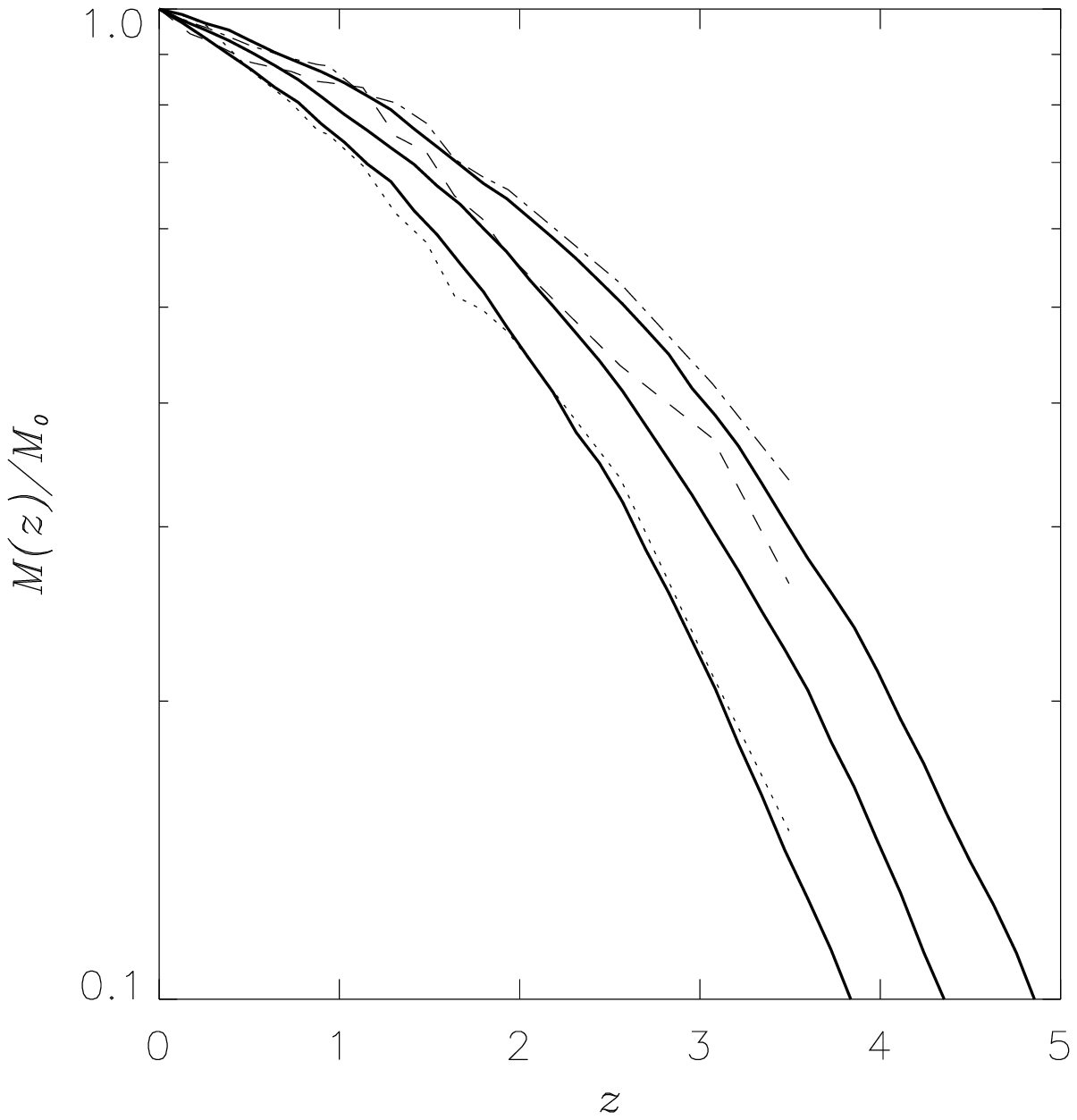}{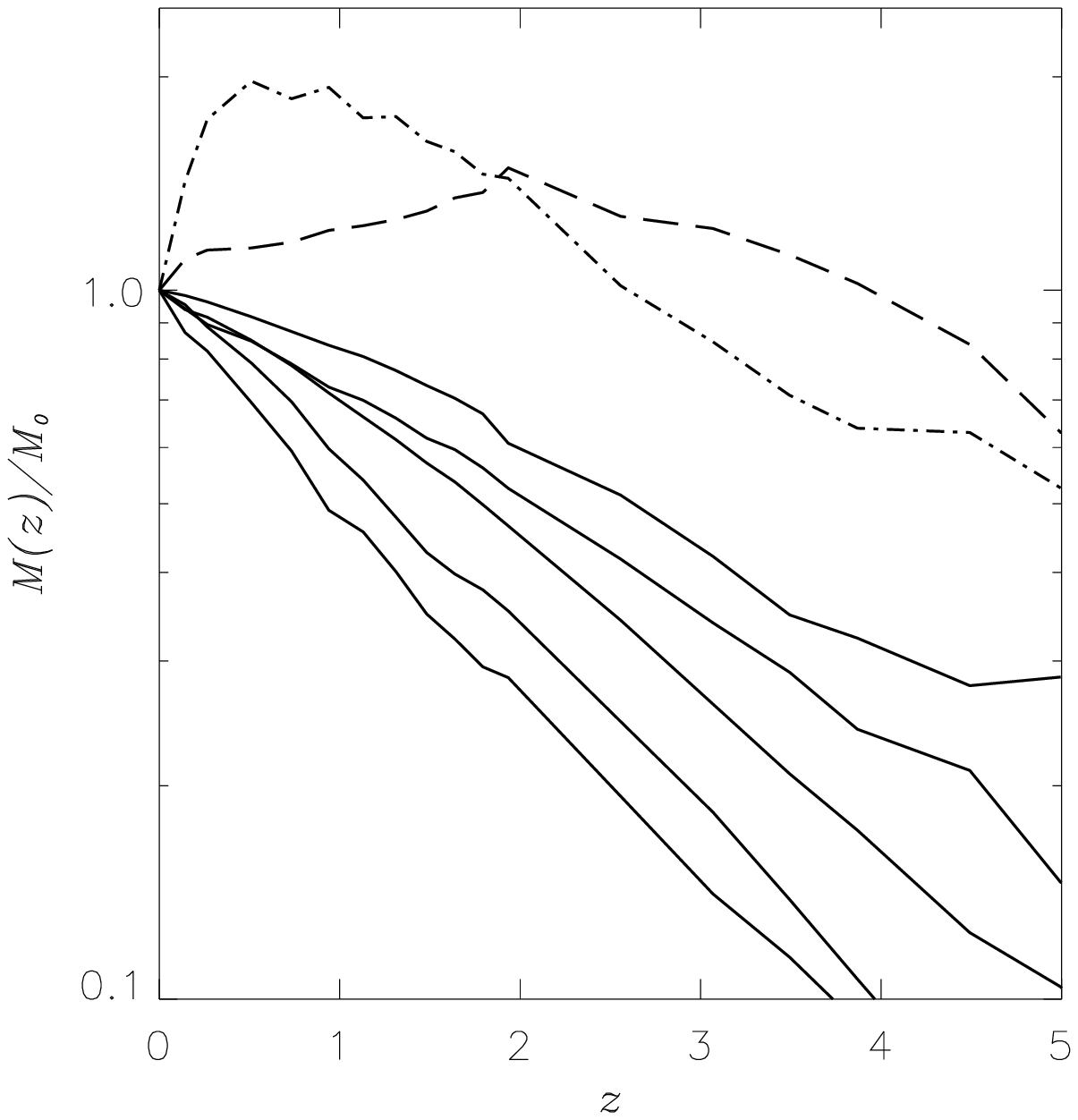}
\caption{Left: Mass evolution of isolated objects identified by the
HFOF algorithm at virial overdensity. The thin lines are for different
average masses of these objects (see text).  The solid lines show the
predictions of the extended Press-Schechter approximation for different
masses (see text).  
Right: Mass evolution of all halos. The solid lines are for different
average masses (see text), the dot-dashed and the dashed lines show the
mass evolution of a subset of halos which loose mass due to the tidal
interaction. The radius of the halos was restricted to be not more than
100 $h^{-1}$ kpc.}
\label{halo_evol}
\end{figure}

The mass of an object found by the HFOF algorithm {\em at virial
overdensity} can be defined as the sum of linked particle masses. In
this case we do not only find galaxy size halos but also all group size
and cluster size halos. For all of the HFOF objects we identify the
main progenitors at all epochs down to the halo detection time. To
study the mass evolution due to merging and accretion we have divided
these objects into three mass bins at $z=0$. The average mass evolution
of halos in these bins normalized to the mass at $z = 0$ is shown in
Figure~\ref{halo_evol}, left. The dotted line is for average halo mass of
$1.2 \times 10^{13}{\rm M_{\odot}}$ (12 halos), the dashed for $1.0
\times 10^{12}{\rm M_{\odot}}$ (17 halos), the dash-dotted for $1.1
\times 10^{11}{\rm M_{\odot}}$ (265 halos). The solid
lines show the predictions of semi-analytical model (kindly provided by
Claudio Firmani) for $10^{13}$, $10^{12}$, and $10^{11} {\rm
M_{\odot}}$, from the bottom to the top. We find a good agreement with
the semi-analytical predictions (Lacey \& Cole 1993) for the evolution
of the FOF selected objects, which are per definition isolated.

Unfortunately, there is no simple and straightforward way to assign a
mass for all halos identified in the simulation. Unlike the isolated
halos identified by HFOF at virial overdensity, the satellite halos,
although surviving, are subject to tidal stripping which reduces
their mass. They are limited therefore by tidal, rather then virial,
radius. To assign masses to the halos we proceed as follows. The
isolated halos are assigned the mass inside the virial radius or radius
of $100 h^{-1}$ kpc, whichever is smaller.  The satellite halos are
assigned the total mass of {\em gravitationally bound} particles within
their tidal radius (or, again, within $100 h^{-1}$ kpc, whichever is
smaller). The tidal radius is determined as the radius at which the
density profile of a halo flattens (stops decreasing).

We now construct the complete mass evolution histories for the set of
{\em all} halos with the masses assigned as described above.  We have
divided these halos into five groups with masses $M_0 > 10^{13}{\rm
M_{\odot}}$, $ 10^{13}{\rm M_{\odot}} > M_0 > 5 \times 10^{12}{\rm
M_{\odot}}$, $ 5 \times 10^{12}{\rm M_{\odot}} > M_0 > 10^{12}{\rm
M_{\odot}}$, $ 10^{12}{\rm M_{\odot}} > M_0 > 5 \times 10^{11}{\rm
M_{\odot}}$, and $ 5 \times 10^{11}{\rm M_{\odot}} > M_0$. We defined a
subset of 3674 halos, mass of which increases (with allowance for small
statistical fluctuations) at all epochs.  As before, the mass of these
objects is normalized to their final mass at $z=0$.  The mass evolution
of these halos is shown in Figure~\ref{halo_evol}, right (solid
lines). The solid lines are for average masses of (from the bottom to
the top) $1.2 \times 10^{13}{\rm M_{\odot}}$ (14 halos), $6.6 \times
10^{12}{\rm M_{\odot}}$ (34 halos), $1.9 \times 10^{12}{\rm M_{\odot}}$
(442 halos), $7.0 \times 10^{11}{\rm M_{\odot}}$ (534 halos), and $2.4
\times 10^{11}{\rm M_{\odot}}$ (2650 halos). The overall
evolution is similar to the mass evolution of isolated halos described
above Figure~\ref{halo_evol}, left). Note, however, that while the mass
evolution tracks are curved in Figure~\ref{halo_evol}, left, the mass
evolution of the sample that includes satellites can be better
represented by the straight lines in these log-log plots. This
difference is due to the different halo selection procedure and to the
different assignment of mass to the selected halos.

In the two lowest mass ranges we also find an additional subset of 2650
halos, whose masses decrease after $z=1$: The dot-dashed (average mass
of $6.9 \times 10^{11}{\rm M_{\odot}}$) and the dashed (average mass of
$2.0 \times 10^{11}{\rm M_{\odot}}$) lines in Figure~\ref{halo_evol}
(right) show the mass evolution of this subset of halos which loose mass
due to the tidal stripping in groups and clusters.

\begin{figure}
\plottwo{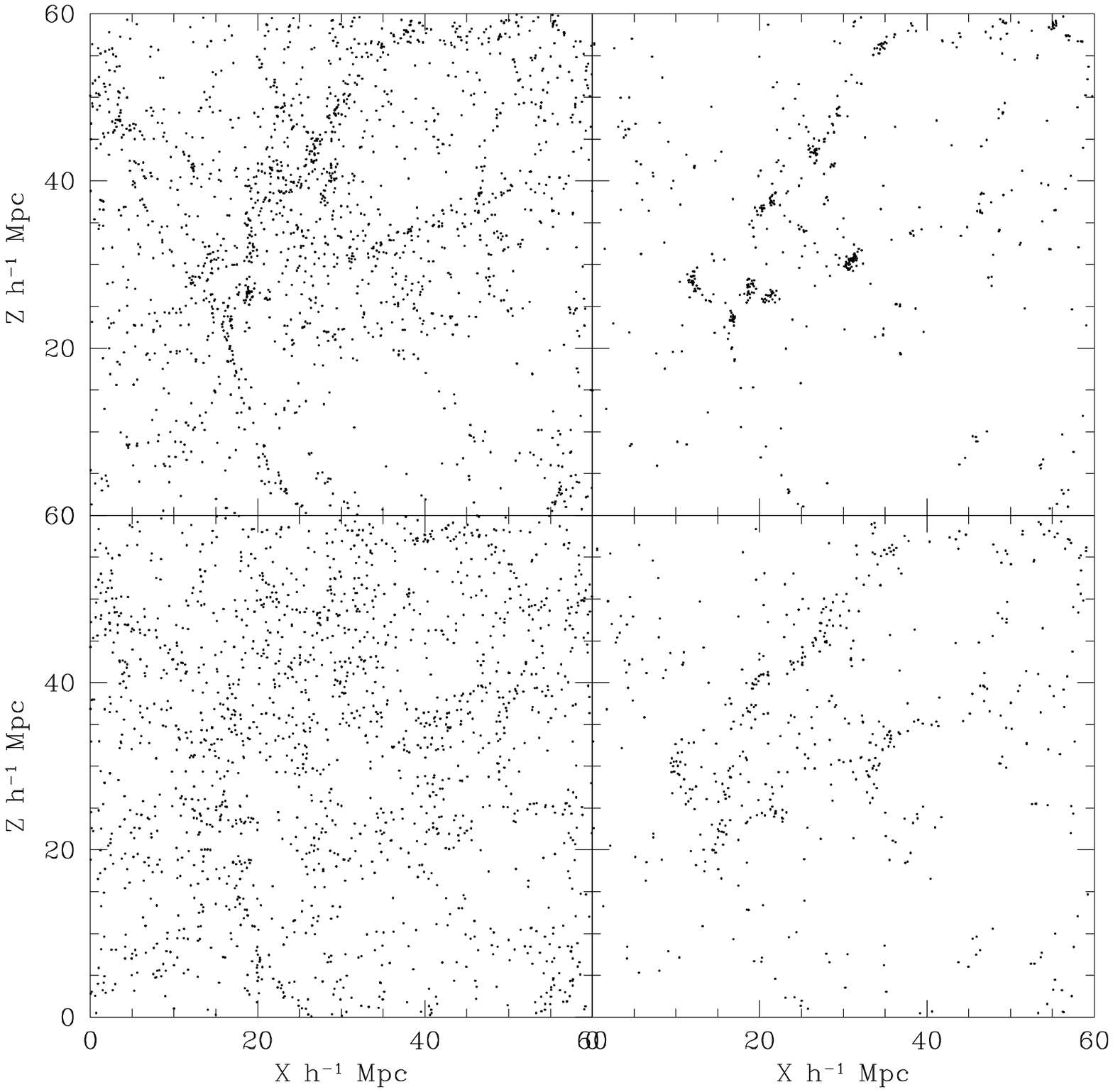}{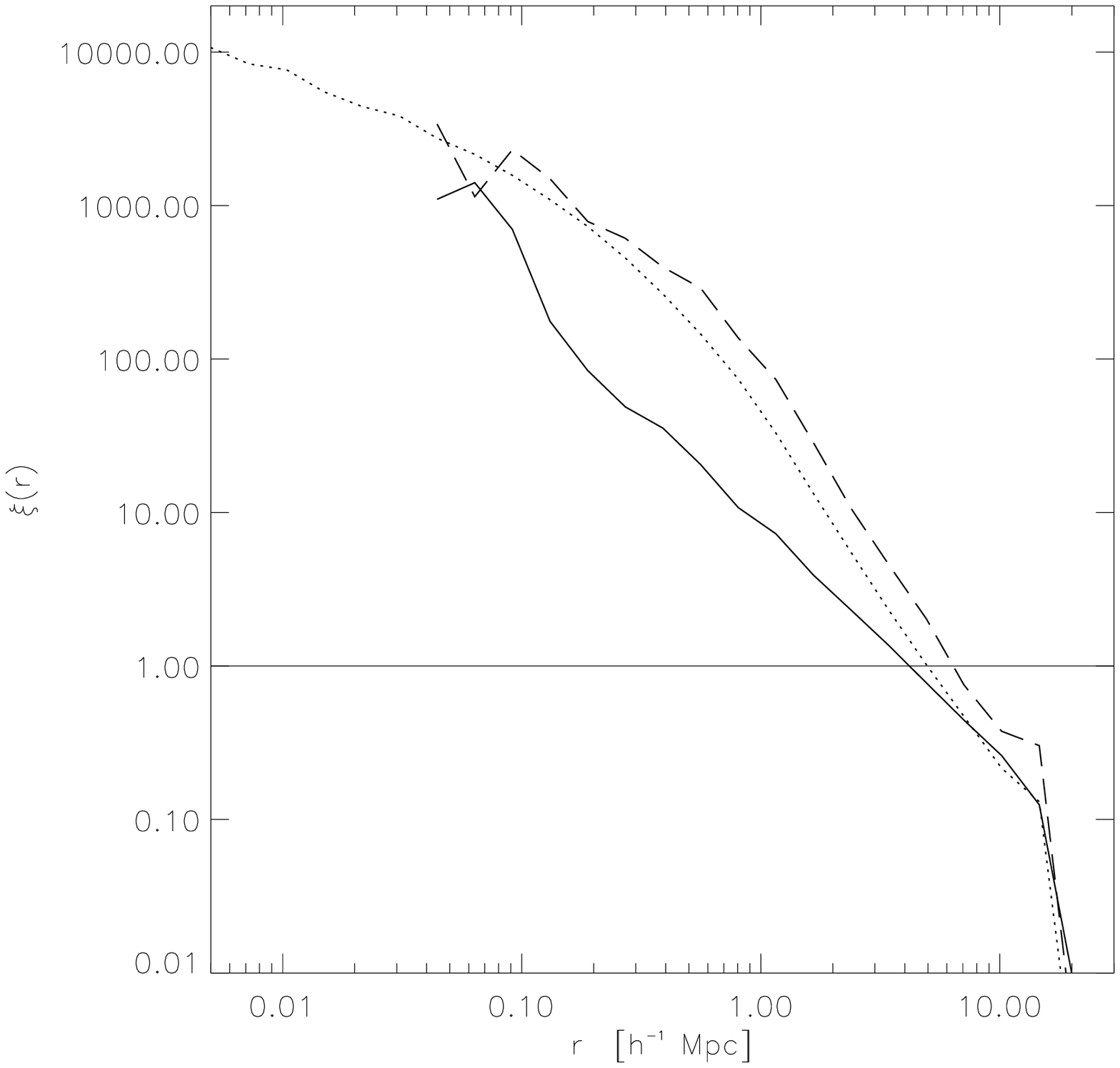}
\caption{Left: A slice of $15 h^{-1}$ Mpc thickness: Merging (left) 
and stripped (right) halos at $z = 0$ (top row), the same halos at $z=1$
(bottom row). Right: The correlation function of halos with circular
velocities $v_{circ} > 120$ km/s at $z=0$. The solid line corresponds to halos
the mass of which always increases (solid lines in
Figure~\ref{halo_evol}, right), the dashed line corresponds to halos
which loose mass during evolution (dashed and dashed-dotted lines in
Figure~\ref{halo_evol}, right).  The correlation function of dark
matter particles is shown by the dotted line.}
\label{corr_fun}
\end{figure}

Their mass increases at high redshifts, reaches a
maximum and decreases thereafter. Contrary to the main halo
population, the mass of which always increases ({\it merging halos})
the mass of these {\it stripped} halos grows first due to accretion of
surrounding material and of smaller halos.  At some point, however,
these halos start to loose mass due to either tidal stripping or
interaction with other nearby halos. In the left panel of
Figure~\ref{corr_fun}, we show the distribution of these halo subsets
at $z=1$ and $z=0$. At $z=1$ the ``stripped'' halos are distributed
similarly to the rest of the halos. At $z=0$, however, they are
clustered much more strongly than the overall halo population.

In the right panel of Figure~\ref{corr_fun}, we show the correlation function
for the two subsets of halos: always increasing mass and decreasing
mass at $z<1$.  The correlation functions of the former has a lower
amplitude and is not as steep as the correlation function of the
latter.  Note that the CF of the halos with the ever-increasing mass is
anti-biased at scales ${_ <\atop{^\sim}} 10 h^{-1}$ Mpc, the CF of the
halos that loose mass is actually positively biased. This reflects the
fact that the loosing mass halos are found within massive systems such
as massive galaxies, groups, and clusters, and are therefore strongly
clustered.

One might speculate that this difference in the correlation functions may
serve as a possible explanation for the color segregation of the
correlation amplitude that has been recently observed (Carlberg et al.
1998). In fact, one could expect that the galaxies hosting halos which
undergo different mass evolution also show different properties, and
colors in particular. Further studies are necessary to test whether
this simple picture can really explain the observations.

\acknowledgments
This work was funded by the NSF and NASA grants to NMSU.  SG
acknowledges support from Deutsche Akademie der Naturforscher
Leopoldina with means of the Bundesministerium f\"ur Bildung und
Forschung grant LPD 1996.  We acknowledge support by NATO grant CRG
972148.  We thank Claudio Firmani and Vladimir Avila-Reese for
providing mass evolution predictions of semi-analytical models. The
numerical simulations has been carried out at the Origin2000 computers
at NCSA and Naval Research Laboratory (NRL).

\end{document}